\documentclass[conference]{IEEEtran}
\IEEEoverridecommandlockouts
\usepackage{
amsmath,cite,bm,amsfonts,amssymb,
graphicx,color,mathtools,setspace,
comment,multirow,xfrac,epstopdf,
footnote,multirow,lscape,float,
caption,subcaption,amsthm,alltt,placeins,amsthm}
\usepackage[ruled,vlined]{algorithm2e}

\newcommand{\myincludegraphics}{\includegraphics[width=1\columnwidth]}

\newcommand{\myVM}[3]{\mathbf{#1}_{\mathrm{#2}}^{#3}} 
\newcommand{\myVMIndex}[4]{\mathbf{#1}_{\mathrm{#2},#3}^{#4}} 
\newcommand{\Norm}[3]{\left\lVert #1 \right\rVert_{#2}^{#3}} 

\newcommand{\eq}[1]{(#1)}
\newcommand{\Diag}[1]{\mathrm{diag}\left\{#1\right\}}
\newcommand{\Det}[1]{\left\lvert #1 \right\rvert}
\newcommand{\Abs}[2]{\left\lvert #1 \right\rvert^{#2}}
\newcommand{\Exp}[1]{\mathbb{E}\left\{ #1 \right\}}

\newcommand{\kth}[1]{#1^{\mathrm{th}}}

\newcommand{\betabr}[0]{\beta_{\mathrm{br}}}

\newcommand{\etad}[0]{\eta_{\mathrm{d}}}
\newcommand{\zetad}[0]{\zeta_{\mathrm{d}}}
\newcommand{\etaru}[0]{\eta_{\mathrm{ru}}}
\newcommand{\zetaru}[0]{\zeta_{\mathrm{ru}}}
\newcommand{\etabr}[0]{\eta_{\mathrm{br}}}
\newcommand{\zetabr}[0]{\zeta_{\mathrm{br}}}

\newcommand{\kd}[0]{\kappa_{\mathrm{d}}}
\newcommand{\kru}[0]{\kappa_{\mathrm{ru}}}
\newcommand{\kbr}[0]{\kappa_{\mathrm{br}}}

\newcommand{\Hd}[1]{\myVM{H}{d}{#1}}
\newcommand{\Hru}[1]{\myVM{H}{ru}{#1}}
\newcommand{\Hbr}[1]{\myVM{H}{br}{#1}}
\newcommand{\Ubr}[1]{\myVM{U}{}{#1}}

\newcommand{\wwr}[1]{\myVM{w}{}{#1}}

\newcommand{\ab}[1]{\myVM{a}{b}{#1}}
\newcommand{\ar}[1]{\myVM{a}{r}{#1}}

\newcommand{\argmax}[1]{\underset{#1}{\text{argmax}}}
\newcommand{\maxover}[1]{\underset{#1}{\text{max}}}

\begin{document}

\bstctlcite{IEEEexample:BSTcontrol} 

\title{Tight Bounds on the Optimal UL Sum-Rate of MU RIS-aided Wireless Systems}

\author{\IEEEauthorblockN{%
		Ikram Singh\IEEEauthorrefmark{1}, %
		Peter J. Smith\IEEEauthorrefmark{2}, %
		Pawel A. Dmochowski\IEEEauthorrefmark{1}}
	\IEEEauthorblockA{\IEEEauthorrefmark{1}%
		School of Engineering and Computer Science, Victoria University of Wellington, Wellington, New Zealand}
	\IEEEauthorblockA{\IEEEauthorrefmark{2}%
		School of Mathematics and Statistics, Victoria University of Wellington, Wellington, New Zealand}
	\IEEEauthorblockA{email:%
		~\{ikram.singh,peter.smith,pawel.dmochowski\}@ecs.vuw.ac.nz
	}%
}

\maketitle
\begin{abstract}
The objective of this paper is to develop simple techniques to bound the optimal uplink sum-rate of multi-user RIS-aided wireless systems. Specifically, we develop a novel technique called \textit{channel separation} which provides a new  understanding as to how the RIS phases affect the sum-rate. Leveraging channel separation, we derive upper and lower bounds on the optimal sum-rate. In addition, we propose a low-complexity alternating optimization algorithm to obtain near-optimal sum-rate results. Numerical results demonstrate the tightness of the bounds and show that the alternating optimization approach delivers sum-rate values similar to the results of a full numerical optimization procedure. Furthermore, in practical scenarios where hardware limitations cause the RIS phases to be quantized, our lower bound can still be applied and shows that the sum-rate is robust to quantization, even with low resolution.
\end{abstract}
\IEEEpeerreviewmaketitle
%
%
\section{Introduction}
Reconfigurable Intelligent Surface (RIS) technology is designed to manipulate the channel between users (UEs) and base station (BS) via the RIS in a wireless system \cite{8796365}. Assuming that channel state information (CSI) is known, then it is possible to intelligently configure the RIS phases to enhance performance metrics (e.g. energy efficiency \cite{8741198}). However, it has become apparent  that  the unit modulus constraint introduced by the RIS phases leads to difficult, non-convex optimization problems for most system metrics  \cite{9090356}. 

A very common optimization problem for wireless communication systems is to maximum the sum-rate among multiple users  \cite{9110889,9286726,9203956,9090356}. In \cite{9110889} an iterative algorithm is proposed to maximize the sum-rate in the absence of a direct channel between the BS and users. Specifically, the DL sum-rate is maximized subject to discrete phases at the RIS and ZF beamforming at the BS given a maximum power threshold. The results presented show that good sum-rate performance can be achieved by a  RIS of appropriate size along with low resolution for the RIS phases. Practically, it is difficult to implement continuous phase control for the RIS, so achieving near optimal performance with low phase resolution is important. The work in \cite{9286726} is similar to that of \cite{9110889} but considers a cell-free environment with multiple RIS deployed to aid transmission from a single BS. In \cite{9203956} the sum-rate is maximized for an UL non-orthogonal multiple access (NOMA) system while successive interference cancellation (SIC) is performed at the single antenna BS in the absence of a UE-BS channel. To find the sub-optimal RIS phases, the authors reformulate the sum-rate maximization problem into the maximization of a quadratic form. The CVX package is then utilized to solve the related semi-definite-relaxation (SDR) problem. In multi-cell environments, \cite{9090356} maximizes the weighted sum-rate of all users through a joint optimization of the precoding matrices at the BSs and of the RIS phases. The authors propose the use of Majorization-Minimization (MM) and Complex Circle Manifold (CCM) methods to optimize the RIS phases whilst keeping the precoding matrices fixed.

Due to the passive nature of the RIS, causing the unit modulus constraint, methods to compute sum-rate are usually algorithmic rather than closed form. Hence, in this paper, we make the following contributions:
\begin{itemize}
    \item We introduce a novel technique called {\textit{channel separation}} which creates an equivalent channel matrix separated into two parts; one part is independent of the RIS and another part consists of a single row directly impacted by the RIS. Channel separation is designed for scenarios where the RIS-BS channel has a strong LOS component.
    \item Leveraging channel separation, the problem of computing the optimal uplink (UL) sum-rate is reformulated to maximizing a simple quadratic form which leads to tight upper and lower bounds on the optimal sum-rate. 
    \item We propose a very low-complexity alternating optimization (AO) algorithm to obtain near optimum results.
    \item Numerical results demonstrate the effectiveness of our techniques whilst showing that quantization of the  RIS phases leads to little sum-rate degradation even with low resolution for the RIS phases.
\end{itemize}

\textit{Notation:} 
$\Norm{\cdot}{1}{}$ denotes the 
$\ell_{1}$ norm. The transpose and Hermitian transpose  are denoted as $(\cdot)^{T}$ and $(\cdot)^{H}$ respectively. The angle of a complex number, $z$, is denoted  $\angle z$. 
 The Kronecker product is denoted $\otimes$. $\mathcal{U}(a,b)$ denotes a uniform random variable taking on values between $a$ and $b$, $\mathcal{N}(\mu,\sigma^2)$ denotes a Normal distribution with mean $\mu$ and variance $\sigma^2$ and $\mathcal{L}(1/{\sigma})$ denotes a Laplacian distribution with standard deviation parameter $\sigma$.

\section{Channel and System Model}\label{Sec: Channel Model}
\begin{figure}[h]
	\centering
	\includegraphics[width=8cm]{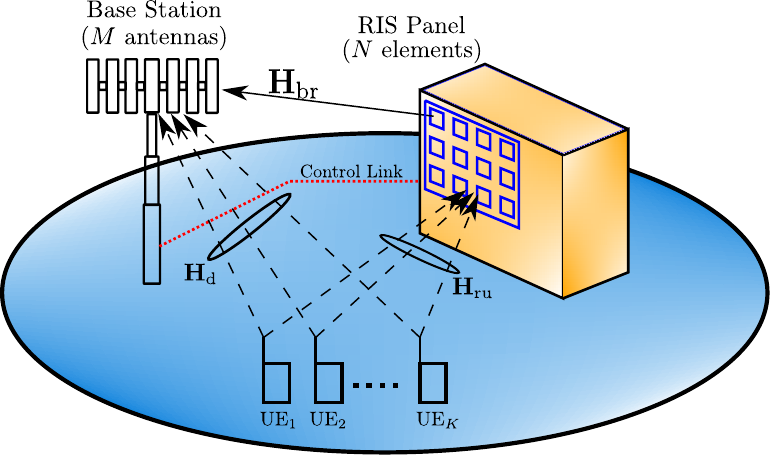}
	\caption{System model.}
	\label{Fig: System Model}
\end{figure}

As shown in Fig.~\ref{Fig: System Model}, we examine a RIS-aided wireless system where a RIS with $N$ reflective elements supports  UL transmission between $K$ single antenna UEs and a BS with $M$ antennas. 

Let $\myVM{H}{d}{} \in \mathbb{C}^{M \times K}$, $\myVM{H}{ru}{} \in \mathbb{C}^{N \times K}$, $\myVM{H}{br}{} \in \mathbb{C}^{M \times N}$ be the UE-BS, UE-RIS, RIS-BS channels, respectively. The diagonal matrix $\myVM{\Phi}{}{} \in \mathbb{C}^{N \times N}$, where $\mathbf{\Phi}_{rr} = e^{j\phi_r}$ for $r=1,2,\ldots,N$, contains the reflection coefficients for each RIS element. Given these matrices, the global UL channel is given by,
\begin{equation}\label{Eq: Global Channel}
	\myVM{H}{}{} = \myVM{H}{d}{} + \myVM{H}{br}{}\mathbf{\Phi}\myVM{H}{ru}{}.
\end{equation}
In the channel model, we adopt a LOS version of the clustered, ray-based model in \cite{8812955} for $\myVM{H}{d}{}, \myVM{H}{ru}{}$:
\begin{equation}\label{Eq: Channels}
	\begin{split}
		\myVM{H}{d}{} & =  \etad \myVM{A}{d}{\mathrm{LOS}}\myVM{B}{d}{1/2} + \zetad\sum_{c=1}^{C_{\mathrm{d}}}\sum_{s=1}^{S_{\mathrm{d}}}\myVM{A}{d,c,s}{\mathrm{SC}}, \\
		\myVM{H}{ru}{} & = \etaru \myVM{A}{ru}{\mathrm{LOS}}\myVM{B}{ru}{1/2} + \zetaru\sum_{c=1}^{C_{\mathrm{ru}}}\sum_{s=1}^{S_{\mathrm{ru}}}\myVM{A}{ru,c,s}{\mathrm{SC}},
	\end{split}
\end{equation}
with 
\begin{align*}
	\etad = \sqrt{\frac{\kd}{1+\kd}}, \quad \zetad = \sqrt{\frac{1}{1+\kd}}, \\
	\etaru = \sqrt{\frac{\kru}{1+\kru}}, \quad \zetaru = \sqrt{\frac{1}{1+\kru}},
\end{align*}
where $C_{\mathrm{d}},C_{\mathrm{ru}}$ are the number of clusters in the UE-BS, UE-RIS channels and $S_{\mathrm{d}},S_{\mathrm{ru}}$ are the number of sub-rays per cluster in the UE-BS and UE-RIS channels. In \eq{\ref{Eq: Channels}}, $\kd$ and $\kru$ are the equivalent of Ricean K-factors for the UE-BS and UE-RIS channels respectively, controlling the relative power of the scattered (ray-based) components and the LOS ray. For simplicity, we assume that each user has the same K-factor, but this can easily be generalized. $\myVM{B}{d}{},\myVM{B}{ru}{}$ are diagonal matrices containing the path gains between UE-BS and UE-RIS respectively, which are modeled by log-normal shadowing. In particular
\begin{equation}\label{Eq: Path gain matrices for Hd and Hru}
    \left( \myVM{B}{d}{} \right)_{kk} = PX_{\mathrm{d},k}d_{\mathrm{d},k}^{-\gamma_{\mathrm{d}}},
    \quad
    \left( \myVM{B}{ru}{} \right)_{kk} = PX_{\mathrm{ru},k}d_{\mathrm{ru},k}^{-\gamma_{\mathrm{ru}}},
\end{equation}
where $d_{\mathrm{d},k}$ and $d_{\mathrm{ru},k}$ are the distances between the $k^{\mathrm{th}}$ UE and the BS and the $k^{\mathrm{th}}$ UE and the RIS respectively, $\gamma_{\mathrm{d}}$ and $\gamma_{\mathrm{ru}}$ are the pathloss exponents, $X_{\mathrm{d},k}$ and $X_{\mathrm{ru},k}$ model the effects of shadow fading and are taken from a log-normal distribution with zero mean and variances $\sigma_{\mathrm{\mathrm{d},sf}}^2$ and $\sigma_{\mathrm{\mathrm{ru},sf}}^2$ respectively. $P$ is the received power at a reference distance of 1m.

$\myVM{A}{d}{\mathrm{LOS}}$ and $\myVM{A}{ru}{\mathrm{LOS}}$ are the LOS components for the UE-BS and UE-RIS channels respectively. The $\kth{k}$ columns of the LOS components for $\Hd{}$ and $\Hru{}$ are given by
\begin{equation}\label{Eq: LOS Component for Hd and Hru}
    \myVM{a}{\mathrm{d},k}{\mathrm{LOS}} = \mathbf{a}_{\mathrm{b}}(\theta_{\mathrm{d}}^{(k)},\phi_{\mathrm{d}}^{(k)}), \
    \myVM{a}{\mathrm{ru},k}{\mathrm{LOS}} = \mathbf{a}_{\mathrm{r}}(\theta_{\mathrm{ru}}^{(k)},\phi_{\mathrm{ru}}^{(k)}),
\end{equation}
where $\theta_{\mathrm{d}}^{(k)}$, $\theta_{\mathrm{ru}}^{(k)}$ are the elevation angles of arrival (AOAs) for the $\kth{k}$ UE and $\phi_{\mathrm{d}}^{(k)}$, $\phi_{\mathrm{ru}}^{(k)}$ are the azimuth AOAs for the $\kth{k}$ UE. Note that the steering vectors at the BS, $\ab{}(\cdot,\cdot)$, and at the RIS, $\ar{}(\cdot,\cdot)$, are topology dependent. Further details are given in Sec.~\ref{Sec: Results}.

$\myVM{A}{d,s,c}{\mathrm{SC}}$ and $\myVM{A}{ru,s,c}{\mathrm{SC}}$ are the scattered components due to the $s$-th subray in the $c$-th cluster which are modeled as in \cite{8812955}. The $k^{\mathrm{th}}$ columns of $\myVM{A}{d,s,c}{\mathrm{SC}}$ and $\myVM{A}{ru,s,c}{\mathrm{SC}}$ are given by the weighted steering vectors,
\begin{equation}\label{Eq: Scattered component for Hd and Hru}
\begin{split}
    \myVMIndex{a}{d,s,c}{k}{\mathrm{SC}} &= \gamma_{\mathrm{d},c,s}^{(k)}\myVM{a}{\mathrm{b}}{}(\theta_{\mathrm{d},c,s}^{(k)},\phi_{\mathrm{d},c,s}^{(k)}), \\
    \myVMIndex{a}{ru,s,c}{k}{\mathrm{SC}} &= \gamma_{\mathrm{ru},c,s}^{(k)}\myVM{a}{\mathrm{r}}{}(\theta_{\mathrm{ru},c,s}^{(k)},\phi_{\mathrm{ru},c,s}^{(k)}),
\end{split}
\end{equation}
where $\theta_{\mathrm{d},c,s}^{(k)}$, $\theta_{\mathrm{ru},c,s}^{(k)}$ are the elevation AOAs and $\phi_{\mathrm{d},c,s}^{(k)}$, $\phi_{\mathrm{ru},c,s}^{(k)}$ are the azimuth AOAs experienced by the $\kth{k}$ UE. The elevation AOAs are calculated by $\theta_{\mathrm{d},c,s}^{(k)} = \theta_{\mathrm{d},c}^{(k)} + \delta_{\mathrm{d},c,s}^{(k)}$ and $\theta_{\mathrm{ru},c,s}^{(k)} = \theta_{\mathrm{ru},c}^{(k)} + \delta_{\mathrm{ru},c,s}^{(k)}$ where $\theta_{\mathrm{d},c}^{(k)},\theta_{\mathrm{ru},c}^{(k)}$ are the central angles for the subrays in cluster $c$ and the deviations of the subrays from the central angle are $\delta_{\mathrm{d},c,s}^{(k)},\delta_{\mathrm{ru},c,s}^{(k)}$. The azimuth AOAs for each ray are $\phi_{\mathrm{d},c,s}^{(k)} = \phi_{\mathrm{d},c}^{(k)} + \Delta_{\mathrm{d},c,s}^{(k)}$ and $\phi_{\mathrm{ru},c,s}^{(k)} = \phi_{\mathrm{ru},c}^{(k)} + \Delta_{\mathrm{ru},c,s}^{(k)}$ where $\phi_{\mathrm{d},c}^{(k)},\phi_{\mathrm{ru},c}^{(k)}$ are the central angles for the subrays in cluster
$c$ and the deviations of the subrays from the central angle are $\Delta_{\mathrm{d},c,s}^{(k)}, \Delta_{\mathrm{ru},c,s}^{(k)}$. $\gamma_{\mathrm{d},c,s}^{(k)}=\beta_{\mathrm{d},c,s}^{(k)1/2}e^{j\psi_{\mathrm{d},c,s}^{(k)}}$ and $\gamma_{\mathrm{ru},c,s}^{(k)}=\beta_{\mathrm{ru},c,s}^{(k)1/2}e^{j\psi_{\mathrm{ru},c,s}^{(k)}}$ are the ray coefficients where the random phases satisfy  $\psi_{\mathrm{d},c,s}^{(k)},\psi_{\mathrm{ru},c,s}^{(k)} \sim
\mathcal{U}(0,2\pi)$ and the ray powers $\beta_{\mathrm{d},c,s}^{(k)}$ and $\beta_{\mathrm{ru},c,s}^{(k)}$ are selected to satisfy $\left(\mathbf{B}_{\mathrm{d}}\right)_{kk} =
\sum_{c=1}^{C_{\mathrm{d}}}\sum_{s=1}^{S_{\mathrm{d}}} \beta_{\mathrm{d},c,s}^{(k)}$ and
$\left(\mathbf{B}_{\mathrm{ru}}\right)_{kk} = \sum_{c=1}^{C_{\mathrm{ru}}}\sum_{s=1}^{S_{\mathrm{ru}}} \beta_{\mathrm{ru},c,s}^{(k)}$.

The majority of the results in this paper are for a pur LOS RIS-BS channel. However, we also show numerically that the results can be applied to scenarios where  $\Hbr{}$ has a smaller scattered component and a dominant LOS path. Hence, we consider the following channel models: 
\subsubsection{$\Hbr{}$ is pure LOS}
\begin{equation}\label{Eq: Hbr pure LOS}
	\Hbr{} = \sqrt{\betabr}\myVM{A}{br}{\mathrm{LOS}},
\end{equation}
with
\begin{equation}\label{Eq: LOS Component for Hbr}
    \myVM{A}{br}{\mathrm{LOS}} = \mathbf{a}_{\mathrm{b}}(\theta_{\mathrm{br,A}},\phi_{\mathrm{br,A}})
    \mathbf{a}_{\mathrm{r}}^{H}(\theta_{\mathrm{br,D}},\phi_{\mathrm{br,D}}),
\end{equation}
where $\theta_{\mathrm{br,A}}$, $\phi_{\mathrm{br,A}}$ are the elevation and azimuth angles of arrival (AOAs) and $\theta_{\mathrm{br,D}}$, $\phi_{\mathrm{br,D}}$ are the elevation and azimuth angles of departure (AODs), $\betabr$ is the link gain between RIS and BS. Here, $\Hbr{}$ is rank-1 and the path gain is $\betabr = d_{br}^{-2}$, where $d_{\mathrm{br}}$ is the distance between RIS-BS.

\subsubsection{$\Hbr{}$ is dominant LOS}
\begin{equation}\label{Eq: Hbr is dominant LOS}
	\myVM{H}{br}{} = \etabr \sqrt{\betabr} \myVM{A}{br}{\mathrm{LOS}} + \zetabr\sum_{c=1}^{C_{\mathrm{br}}}\sum_{s=1}^{S_{\mathrm{br}}}\myVM{A}{br,c,s}{\mathrm{SC}},
\end{equation}
such that $\etabr >> \zetabr$, with
\begin{align*}
	\etabr = \sqrt{\frac{\kbr}{1+\kbr}}, \quad \zetabr = \sqrt{\frac{1}{1+\kbr}}, 
\end{align*}
where $\betabr$ is the path gain between RIS-BS given by $\betabr = d_{\mathrm{br}}^{-2}/\etabr^2$ and $\myVM{A}{br}{\mathrm{LOS}}$  is given by \eq{\ref{Eq: LOS Component for Hbr}}. The $\myVM{A}{br,c,s}{\mathrm{SC}}$ matrices contain the scattered rays and are calculated in the same manner as for the other channels. $\kbr$ is the Ricean K-factor for the RIS-BS channel. 
In scenarios where the BS and RIS are located in close proximity, it is reasonable to assume that the RIS-BS channel is dominated by its LOS component \cite{9066923}.

Using \eq{\ref{Eq: Global Channel}} and the channels described above, the received signal at the BS is, 
\begin{equation}\label{Eq: r}
	\mathbf{r} = \mathbf{H}\mathbf{s} + \mathbf{n},
\end{equation}
where $\mathbf{s}$ is a $K \times 1$ vector of transmitted symbols, each with a power of $\Exp{\Abs{s_k}{2}} = E_{s}$ and $\mathbf{n} \sim \mathcal{CN}(0,\sigma^2\mathbf{I}_{M})$. For our results, we will assume that $E_s=1$ and $\sigma^2=1$.

\section{Channel Separation}\label{Sec: Channel Separation}

The optimal UL sum-rate, $R_{\mathrm{sum}}^{\text{opt}}$, for the system in \eq{\ref{Eq: r}} is obtained by maximizing the traditional sum-rate expression for an UL MU-MIMO channel \cite{dush} over the possible RIS phases in $\Phi$. Hence, we have
\begin{equation}\label{Eq: SR}
         R_{\mathrm{sum}}^{\text{opt}} = \maxover{\mathbf{\Phi}} \quad \log_2\left\lvert \mathbf{I}_{K} + {\mathbf{H}}^H{\mathbf{H}} \right\rvert.
    \end{equation}
Finding the optimal RIS phases to maximize the sum-rate is the associated design problem given by
\begin{equation}\label{Eq: SR_phases}
    \begin{aligned}
        \mathbf{\Phi}^{\text{opt}} =\argmax{\mathbf{\Phi}} \quad  \log_2\left\lvert \mathbf{I}_{K} + {\mathbf{H}}^H{\mathbf{H}} \right\rvert, \\
    \end{aligned}
    \tag{P.1}
\end{equation}
where the maximization is constrained over the unit amplitude diagonal entries in $\mathbf{\Phi}$. The difficulty in finding $\mathbf{\Phi}^{\text{opt}}$ is largely due to the fact that  $\Phi$ affects every element of ${\mathbf{H}}$. Hence, the log-determinant to be maximized is a very complex function of $\Phi$. However, when the RIS-BS link is LOS then $\Hbr{}$ is rank 1 and the RIS phases only affect a rank 1 component of ${\mathbf{H}}$. Motivated by this observation, we seek to separate out this RIS-dependent, rank 1 component from the rest of the channel. We refer to this method as {\textit{channel separation}} and in this section, we assume that the RIS-BS link is pure LOS.

Channel separation is achieved via a unitary transformation of ${\mathbf{H}}$. For any $N \times N$ unitary matrix, ${\mathbf{U}}$, we can define ${\tilde{\mathbf{H}}}={\mathbf{U}}^H{\mathbf{H}}$ and ${\tilde{\mathbf{H}}}^H{\tilde{\mathbf{H}}}={\mathbf{H}}^H{\mathbf{H}}$. Hence, the sum-rate for channel ${\tilde{\mathbf{H}}}$ is identical to the sum-rate with ${\mathbf{H}}$. Substituting the expression for $\Hbr{}$ in \eqref{Eq: Hbr pure LOS} into ${\tilde{\mathbf{H}}}$, we obtain
\begin{equation}\label{CS}
{\tilde{\mathbf{H}}}={\mathbf{U}}^H \Hd{} +\sqrt{\beta_{\textrm{br}}}{\mathbf{U}}^H \ab{} \ar{H} \mathbf{\Phi} \Hru{}.
\end{equation}
Note that $\ab{}$ and $\ar{}$ are used as simplified notation for the steering vectors in \eqref{Eq: LOS Component for Hbr} for the $\Hbr{}$ channel. Since $\ar{H} \mathbf{\Phi} \Hru{}$ is a row vector, we can confine the effects of $\mathbf{\Phi}$ to one row of ${\tilde{\mathbf{H}}}$ by selecting ${\mathbf{U}}$ to satisfy 
\begin{equation}\label{CS_sol}
{\mathbf{U}}^H \ab{} =\frac{1}{\sqrt{N}}[1, 0, \ldots ,0]^T.
\end{equation}
The unitary matrix satisfying \eqref{CS_sol} is the matrix of left singular vectors of $\Hbr{}$ as shown below.

Define the singular value decomposition (SVD) of $\Hbr{}$ as 
$\Hbr{} = \myVM{U}{}{}\myVM{D}{}{}\myVM{V}{}{H}$, 
where $\myVM{U}{}{}=[\myVM{u}{1}{}, \ldots \mathbf{u}_{M}]$ is the matrix of left singular vectors, $\myVM{D}{}{}$ is the diagonal matrix of singular values   and $\myVM{V}{}{}=[\myVM{v}{1}{}, \ldots \mathbf{v}_{N}]$  is the matrix of right singular vectors. Since $\Hbr{}$ is rank-1, then only one non-zero singular value, $d_1$, exists and $\Hbr{} =d_1\myVM{u}{1}{}\myVM{v}{1}{H}$ where   $\myVM{u}{1}{} = \ab{}/\sqrt{M}$, $\myVM{v}{1}{} = \ar{}/\sqrt{N}$ and $d_1 = \sqrt{M N \betabr}$. Using this value of $\myVM{U}{}{}$, we have
\begingroup
\allowdisplaybreaks
\begin{align}
{\tilde{\mathbf{H}}} &=  \Ubr{H}\Hd{} + 
	\begin{bmatrix}
		\ab{H}/\sqrt{M} \\
		\myVM{u}{2}{H} \\
		\vdots \\
		\myVM{u}{M}{H}
	\end{bmatrix}
	\sqrt{\betabr}\ab{}\ar{H}\myVM{\Phi}{}{}\Hru{} \notag \\
	&= 
	\begin{bmatrix}
		\myVM{u}{1}{H}\Hd{} + \sqrt{M\betabr}\ar{H}\myVM{\Phi}{}{}\Hru{} \\
		\myVM{u}{2}{H}\Hd{} \\
		\vdots \\
		\myVM{u}{M}{H}\Hd{}
	\end{bmatrix} \notag \\
	&\triangleq 
	\begin{bmatrix}
		\wwr{H} \\
		\myVM{H}{1}{}
	\end{bmatrix}.\label{chan_sep}
\end{align}
\endgroup
Channel separation is observed in \eqref{chan_sep} where $\wwr{H}$, the first row of ${\tilde{\mathbf{H}}}$, is the only row affected by $\myVM{\Phi}{}{}$.

Using \eqref{chan_sep}, we can express the desired determinant as
\begin{align}\label{QF}
\left\lvert \mathbf{I}_{K} + {\mathbf{H}}^H{\mathbf{H}} \right\rvert & = \left\lvert \mathbf{I}_{K} + {\tilde{\mathbf{H}}}^H{\tilde{\mathbf{H}}} \right\rvert \notag \\
&=\left\lvert \mathbf{I}_{K}  + \myVM{H}{1}{H}\myVM{H}{1}{} + \wwr{}\wwr{H}\right\rvert \notag \\
& \triangleq \left\lvert \mathbf{Q} + \wwr{}\wwr{H}\right\rvert \notag \\
&=\left\lvert \mathbf{Q} \right\rvert \left(1+\wwr{H}\mathbf{Q}^{-1}\wwr{}\right),
\end{align}
where \eqref{QF} follows from the matrix determinant lemma.

In deriving \eqref{QF}, the SVD of the $M \times N$ matrix $\Hbr{}$ was used. However, the final solution can be written in terms of the channels only, making it computationally trivial involving only a $K \times K$ determinant and a $K \times K$ inverse. This is achieved by writing $\myVM{U}{}{}=[\myVM{u}{1}{} \myVM{U}{2}{}]$, so that  $\myVM{U}{}{} \myVM{U}{}{H} =  \myVM{I}{M}{} =\myVM{u}{1}{} \myVM{u}{1}{H}+\myVM{U}{2}{} \myVM{U}{2}{H}$. Using this result gives $\myVM{Q}{}{}=\mathbf{I}_K+\myVM{H}{1}{H}\myVM{H}{1}{} = \mathbf{I}_K+\myVM{H}{d}{H}\myVM{U}{2}{}\myVM{U}{2}{H}\myVM{H}{d}{}=\mathbf{I}_K+\myVM{H}{d}{H}(\myVM{I}{M}{}-\myVM{u}{1}{} \myVM{u}{1}{H})\myVM{H}{d}{}$.

Using \eqref{QF} and noting that $\mathbf{Q}$ is Hermitian, an equivalent statement of the maximization problem in \eq{\ref{Eq: SR_phases}} is
\begin{equation}\label{Eq: sum-rate}
    \begin{aligned}
          \argmax{\mathbf{\Phi}} \quad & 
         \wwr{H}\mathbf{Q}^{-1}\wwr{}
        \\
        \text{s.t.} \quad &\Abs{\Phi_{ii}}{}=1 \text{ for } i=1,\ldots,N.
    \end{aligned}
    \tag{P.2}
\end{equation}
The benefit of channel separation is clearly seen in \eqref{Eq: sum-rate} where maximization is now over a simple scalar quadratic form. Furthermore, substituting $\myVM{u}{1}{} = \ab{}/\sqrt{M}$ into  $\myVM{Q}{}{}$ and $\wwr{}$ gives
\begin{align}
\myVM{Q}{}{} &=\mathbf{I}_K+\myVM{H}{d}{H}\left(\myVM{I}{M}{}-\frac{\myVM{a}{b}{}\myVM{a}{b}{H}}{M}\right)\myVM{H}{d}{},\\
\wwr{} &=\Hd{H}\myVM{a}{b}{}/\sqrt{M} + \sqrt{M\betabr}\Hru{H}\myVM{\Phi}{}{H}\ar{}, \label{Eq: v vector}
\end{align}
so that all terms in  the objective of \eqref{Eq: sum-rate} are simple functions of the channels.

\section{Sum-Rate Maximization}\label{Sec: Sum-Rate Maximization}
In this section, we consider several low-complexity approaches to obtain sub-optimal solutions to the maximization problem \eq{\ref{Eq: sum-rate}} as well as lower and upper bounds to the optimal sum-rate.  Expanding the quadratic form and using \eqref{Eq: v vector} we have
\begin{align}\label{Eq: Quad form expanded}
    &\wwr{H}\mathbf{Q}^{-1}\wwr{} \notag \\
    &=
    \myVM{w}{1}{H}\mathbf{Q}^{-1}\myVM{w}{1}{} + 
    \mathbf{x}^H\mathbf{Z}'\mathbf{Q}^{-1}\mathbf{Z}'^H\mathbf{x} +
    2\Re\{\mathbf{x}^H\mathbf{Z}'\mathbf{Q}^{-1}\myVM{w}{1}{}\},
\end{align}
where $\myVM{w}{1}{} = \Hd{H}\myVM{a}{b}{}/\sqrt{M}$, $\mathbf{Z'} = \sqrt{M\betabr}\Diag{\ar{H}}\Hru{}$ and $\mathbf{x}=[\exp{(-j\phi_1)}, \exp{(-j\phi_2)}, \ldots ,\exp{(-j\phi_N})]^T$ is the vector containing, for ease of exposition, the conjugates of the RIS phase values. Note that the first two terms in \eq{\ref{Eq: Quad form expanded}} are quadratic and dominate the third term. This is further accentuated by any maximizing of the terms over the RIS phases. Hence, as an approximation, we maximize the  dominating terms leading to the maximization of $\mathbf{x}^H\mathbf{Z}\mathbf{x}$
with $\mathbf{Z} = \mathbf{Z'}\mathbf{Q}^{-1}\mathbf{Z'}^H$. Hence, the approximate optimization problem is formulated as
\begin{equation}\label{Eq: P.1}
    \begin{aligned}
          \argmax{\mathbf{x}}  \quad & \mathbf{x}^H \mathbf{Z} \mathbf{x} \\
        \text{s.t.} \quad &\Abs{x_{i}}{}=1 \text{ for } i=1,\ldots,N.
    \end{aligned}
    \tag{P.3}
\end{equation}
Problem \eq{\ref{Eq: P.1}} is a concise statement of the new understanding arising from channel separation. The transformation of the channel in \eqref{chan_sep} resulted in a single row containing the RIS phases. Using \eqref{QF} and  \eqref{Eq: Quad form expanded} this rank-1 component of the channel led to the scalar term $\mathbf{x}^H \mathbf{Z} \mathbf{x}$ which dominates the effect of the RIS phases on sum-rate. Hence, we see that effective RIS phases must align strongly with $\mathbf{Z}$.

Notice that if the constraint in \eq{\ref{Eq: P.1}} is relaxed to $\mathbf{x}^H\mathbf{x}=N$, then the optimum solution, $\mathbf{x}^{\star}$, is proportional to the maximal eigenvector of $\mathbf{Z}$. Direct computation of $\mathbf{x}^{\star}$ requires the eigenvalue decomposition of a $N \times N$ matrix. Alternatively, a  low complexity approach to computing $\mathbf{x}^{\star}$ can be derived as
\begin{equation}\label{maxeig}
	\mathbf{x}^{\star} \propto \mathbf{Z}' \mathbf{x'}^{\star},
\end{equation}
where $\mathbf{x'}^{\star}$ is the  maximal eigenvector of $ \alpha\mathbf{I}_K + \mathbf{Q}^{-1}\mathbf{Z'}^H\mathbf{Z'} $. The problem has been reduced from an $N \times N$ to a $K \times K$ eigenvalue decomposition, a considerable saving especially when considering large RIS sizes. 

\subsection{Lower bound on the optimal sum-rate}\label{SubSec: Closed form approximation}
Due to \eq{\ref{Eq: P.1}} being non-convex, it is very difficult to obtain an exact optimal solution in closed form. As such, we will consider an alternative optimization problem, which is to obtain an approximate solution $\hat{\mathbf{x}}$ to \eq{\ref{Eq: P.1}}. Specifically, we minimize the $\ell_1$-norm of the residuals between $\mathbf{x}^{\star}$ in \eqref{maxeig}, the solution to the relaxed version of \eq{\ref{Eq: P.1}}, and the approximate solution. Mathematically, the alternative optimization problem is
\begin{equation}\label{Eq: P.2}
    \begin{aligned}
        \min \quad & \Norm{\mathbf{x}^{\star} - \hat{\mathbf{x}}}{1}{} = \Abs{x^{\star}_1 - \hat{x}_1}{} + \ldots + \Abs{x^{\star}_N - \hat{x}_N}{}\\
        \text{s.t.} \quad &\Abs{\hat{x}_i}{}=1 \text{ for } i=1,\ldots,N.
    \end{aligned}
    \tag{P.4}
\end{equation}
Problem \eq{\ref{Eq: P.2}} can be solved by minimizing each  residual separately. 
It is straightforward to show that this is achieved by setting $\hat{x}_i = e^{j \angle x^{\star}_i}$ and the approximate solution to \eq{\ref{Eq: P.1}} is
\begin{equation}\label{Eq: Closed form optimal RIS}
    \hat{\mathbf{x}} = [e^{j \angle x^{\star}_1}, \ldots, e^{j \angle x^{\star}_N}].
\end{equation}
Using the phases in \eqref{Eq: Closed form optimal RIS} at the RIS  provides a  lower bound on \eq{\ref{Eq: SR}}.
\subsection{Efficient alternating optimization algorithm}\label{SubSec: Efficient low-complexity iterative algorithm}
Here, we propose a low-complexity AO algorithm to find the RIS phases which maximize the sum-rate. The algorithm iteratively maximizes the $\kth{n}$ RIS reflecting coefficient while keeping the other $N-1$ coefficients fixed. We define the following vectors: $\mathbf{x}_{(-n)} \triangleq [x_{1},\ldots,x_{n-1},0,x_{n+1},\ldots,x_{N}]^T$ and $\mathbf{e}_{n} \triangleq [0,\ldots,0,x_{n},0,\ldots,0]^T$. Let $\myVM{B}{}{} = \mathbf{Z}'\mathbf{Q}^{-1}\mathbf{Z}'^H$, $\myVM{w}{2}{} = \mathbf{Z}'\mathbf{Q}^{-1}\myVM{w}{1}{}$, then by expressing \eq{\ref{Eq: Quad form expanded}} in terms of the $\kth{n}$ reflecting coefficient, we have
\begin{align}\label{Eq: Obj for IA}
    &\wwr{H}\mathbf{Q}^{-1}\wwr{} \notag \\
    &= \left(\myVM{w}{1}{H} + \left(\mathbf{x}_{(-n)}^H + \mathbf{e}_{n}^H\right)\mathbf{Z}' \right)\mathbf{Q}^{-1}\left(\myVM{w}{1}{} + \mathbf{Z}'^H \left(\mathbf{x}_{(-n)} + \mathbf{e}_{n} \right) \right) \notag \\
    &=  T_1+ 2\Re\{ \mathbf{e}_{n}^H  \myVM{B}{}{}  \mathbf{x}_{(-n)} + \mathbf{e}_{n}^H \myVM{w}{2}{} \},\notag \\
    &=  T_1+ 2\Re\{ {x}_{n}^*(  \myVM{b}{n}{H}  \mathbf{x}_{(-n)} +  {w}_{2n}) \} ,
\end{align}
where $T_1$ contains the terms not including $\mathbf{e}_{n}$, $w_{2n}$ is the $n$-th element of $\myVM{w}{2}{}$ and $\myVM{b}{n}{}$ is the $n$-th column of $\myVM{B}{}{}$.

Since the quadratic form in \eqref{Eq: Obj for IA} is positive, it is maximized over $x_n$ by maximizing $2\Re\{ {x}_{n}^*(  \myVM{b}{n}{H}  \mathbf{x}_{(-n)} +  {w}_{2n}) \} $. Hence, in order to optimize the $\kth{n}$ reflecting coefficient while the others coefficients are fixed, we can set
\begin{equation}\label{Eq: ith optimal RIS coefficient}
    x_n^{\textrm{update}} = e^{j\angle \left( \mathbf{b}_{n}^H\mathbf{x}_{(-n)} + w_{2n} \right)}.
\end{equation}
The computations involved in \eqref{Eq: ith optimal RIS coefficient} are trivial as $\myVM{B}{}{}$  and $\myVM{w}{2}{}$ are one-off calculations requiring only a $K \times K$ matrix inverse and matrix multiplications. Using \eq{\ref{Eq: ith optimal RIS coefficient}} as the updating equation for each iteration in the algorithm, the AO algorithm is given in Algorithm~\ref{Algorithm 1}.
\begin{algorithm}[h]
	\SetAlgoLined
	\KwResult{RIS reflection coefficients, $\mathbf{x}^{\star}$}
	Set algorithm precision threshold $\epsilon > 0$ \\
	Set initial RIS coefficients to $\mathbf{x} = \mathbf{1}_N$ \\
	Calculate initial $\mathbf{w}$ using \eq{\ref{Eq: v vector}} where $\mathbf{\Phi}=\Diag{\mathbf{x}}$ \\
	Set $T = 0$ \\
	\While{$(\mathbf{w}^H\mathbf{Q}^{-1}\mathbf{w} - T) \geq \epsilon$}{
	    Calculate $T = \mathbf{w}^H\mathbf{Q}^{-1}\mathbf{w}$ \\
	    \For{$n=1:N$}{
	        Update $\kth{n}$ RIS coefficient using \eq{\ref{Eq: ith optimal RIS coefficient}} \\
	    }
	    Calculate $\mathbf{w}$ using \eq{\ref{Eq: v vector}} where $\mathbf{\Phi}=\Diag{\mathbf{x}}$\\
	}
	Return $\mathbf{x}^{\star} = \mathbf{x}$.
	\caption{Sum-Rate AO Algorithm}
	\label{Algorithm 1}
\end{algorithm}

\subsection{Upper bound on the optimal sum-rate}
Leveraging channel separation, we can also derive an upper bound on the sum-rate when the RIS-BS channel is LOS. Substituting \eqref{Eq: Quad form expanded} into \eqref{QF}, along with the Cauchy–Schwarz inequality to bound the elements of the second and third terms in \eqref{Eq: Quad form expanded}, we have
\begin{align}\label{Eq: Upper bound on Sum-Rate}
    &R_{\mathrm{sum}}  
    \leq
    \log_2\left(\Det{\mathbf{Q}}\right) \notag \\
    & \quad +
    \log_2\left( \myVM{w}{1}{H}\mathbf{Q}^{-1}\myVM{w}{1}{} + \sum_{n=1}^{N}\sum_{r=1}^{N}\Abs{B_{nr}}{} +
    2\sum_{r=1}^{N}\Abs{p_{n}}{}  + 1\right),
\end{align}
where $\mathbf{B} = \mathbf{Z'}\mathbf{Q}^{-1}\mathbf{Z'}^H$ and $\mathbf{p}= \mathbf{Z}'\mathbf{Q}^{-1}\myVM{w}{1}{}$.

\section{Results}\label{Sec: Results}
We now demonstrate the effectiveness of the different techniques presented in Sec.~\ref{Sec: Sum-Rate Maximization}. Users were randomly located in a cell with a radius of 50m, outside an exclusion radius of 5m around the BS and RIS. As stated in Sec.~\ref{Sec: Channel Model}, the steering vectors used in the channels are topology dependent. We assume an $M$-element vertical uniform rectangular array in the $y-z$ plane \cite{8812955}   with equal spacing in both dimensions at both the BS and RIS. The $y$ and $z$ components of a generic steering vector at the BS for a given elevation angle, $\theta$, and azimuth angle, $\phi$, are given by,
\begin{align*}
    \mathbf{a}_{\mathrm{b},y}\left( \theta,\phi \right)
    &= 
    [1,\ldots,e^{j2\pi(M_y-1)d_{\mathrm{b}}\sin(\theta)\sin(\phi)} ]^T, \\
    \mathbf{a}_{\mathrm{b},z}\left( \theta,\phi \right)
    &= 
    [1,\ldots,e^{j2\pi(M_y-1)d_{\mathrm{b}}\cos(\theta)} ]^T,
\end{align*}
respectively, where $M = M_{y}M_{z}$ with $M_{y}, M_{z}$ being the number of antenna columns, rows at the BS and  $d_{\mathrm{b}}=0.5$ is the antenna separation in wavelength units. Similarly at the RIS, we have
\begin{align*}
    \mathbf{a}_{\mathrm{r},y}\left( \theta,\phi \right)
    &= 
    [1,\ldots,e^{j2\pi(M_y-1)d_{\mathrm{r}}\sin(\theta)\sin(\phi)} ]^T, \\
    \mathbf{a}_{\mathrm{r},z}\left( \theta,\phi \right)
    &= 
    [1,\ldots,e^{j2\pi(M_y-1)d_{\mathrm{r}}\cos(\theta)} ]^T,
\end{align*}
respectively where $N = N_{y}N_{z}$ with $N_{y},N_{z}$ being the number of columns, rows of RIS elements, $d_{\mathrm{r}}=0.2$ is the RIS element separation in wavelength units. The generic steering vectors at the BS and RIS are then given by,
\begin{equation}\label{Eq: Steering vectors at BS and RIS}
\begin{split}
    \mathbf{a}_{\mathrm{b}}\left( \theta,\phi \right) &= \mathbf{a}_{\mathrm{b},y}\left( \theta,\phi \right) \otimes \mathbf{a}_{\mathrm{b},z}\left( \theta,\phi \right),
    \\
    \mathbf{a}_{\mathrm{r}}\left( \theta,\phi \right) &= \mathbf{a}_{\mathrm{r},y}\left( \theta,\phi \right) \otimes \mathbf{a}_{\mathrm{r},z}\left( \theta,\phi \right).
\end{split}
\end{equation}
Note that \eq{\ref{Eq: Steering vectors at BS and RIS}} can be used to generate all of the channels in Sec.~\ref{Sec: Channel Model} by substituting the relevant elevation and azimuth angles. 
For the LOS components in channels $\Hd{}$ and $\Hru{}$, the elevation and azimuth AOAs for the $\kth{k}$ UE are generated using $\theta_{\mathrm{d}}^{(k)}, \theta_{\mathrm{ru}}^{(k)}\sim \mathcal{U}(0,\pi)$, $\phi_{\mathrm{d}}^{(k)}, \phi_{\mathrm{ru}}^{(k)} \sim \mathcal{U}(-\pi/2,\pi/2)$. For the LOS component of $\myVM{H}{br}{}$, we assume that the elevation and azimuth angles are selected based on the following geometry representing a range of LOS links with less elevation variation than azimuth variation: 
$ \theta_{D} \sim \mathcal{U}[70^{o},90^{o}] $, $ \phi_{D} \sim \mathcal{U}[-30^{o},30^{o}] $,
$ \theta_{A} = 180^{o} - \theta_{D} $, $ \phi_{A} \sim \mathcal{U}[-30^{o},30^{o}] $. 

For the rays in the scattered components, we model all central and deviation elevation angles by \cite{8812955}: $\theta_{\mathrm{E},c}^{(k)}\sim\mathcal{L}(1/\hat{\sigma}_{\mathrm{E},c}) , \delta_{\mathrm{E},c,s}^{(k)}\sim\mathcal{L}(1/\hat{\sigma}_{\mathrm{E},s})$ and the central and deviation azimuth angles by $\phi_{\mathrm{E},c}^{(k)}\sim\mathcal{N}(\mu_{\mathrm{E},c},\sigma_{\mathrm{E},{c}}^2),  \Delta_{\mathrm{E},c,s}^{(k)}\sim\mathcal{L}(1/\sigma_{\mathrm{E},s})$. The subscript $\mathrm{E} \in \{\mathrm{d},\mathrm{ru},\mathrm{br}\}$ represents the different channels. We assume that the parameter values for generating the subrays for each cluster are identical for both $\Hd{}$ and $\Hru{}$, and represent a broad spread of rays, as given in \cite{8812955}. For channel $\Hbr{}$, we assume that the rays are narrowly spread, for which the parameter values are also given in \cite{8812955}. Specifically, the system parameter values are given in Table \ref{Table 1} for all results unless otherwise specified.
\begin{table}[h]
    \centering
    \begin{tabular}{|c | c  |} 
     \hline
     Parameter & Values \\
     \hline
     Cell Radius & 50 m \\
     Exclusion Radius & 5 m \\
     BS Antennas & 32 \\
     \hline
     \underline{Channels $\Hd{}, \Hru{}$} &\\
     $C_{\mathrm{d}}=C_{\mathrm{ru}}$ & 20 \\
     $S_{\mathrm{d}}=S_{\mathrm{ru}}$ & 20 \\
     $\mu_{\mathrm{d},c}=\mu_{\mathrm{ru},c}$, & 0$^{\circ}$ \\
     $\sigma^2_{\mathrm{d},c}=\sigma^2_{\mathrm{ru},c}, 
     \sigma^2_{\mathrm{d},s}=\sigma^2_{\mathrm{ru},s}$ & 31.64$^{\circ}$, 24.25$^{\circ}$ \\
     $\hat{\sigma}^2_{\mathrm{d},c}=\hat{\sigma}^2_{\mathrm{ru},c},\hat{\sigma}^2_{\mathrm{d},s}=\hat{\sigma}^2_{\mathrm{ru},s}$ & 6.12$^{\circ}$, 1.84$^{\circ}$ \\
     \hline 
     \underline{Channel $\Hbr{}$} & \\
     $C_{\mathrm{br}}$, $S_{\mathrm{br}}$ & 3, 16 \\
     $\mu_{\mathrm{br},c}$ & 0$^{\circ}$ \\
     $\sigma^2_{\mathrm{br},c},
     \sigma^2_{\mathrm{br},s},\hat{\sigma}^2_{\mathrm{br},c},\hat{\sigma}^2_{\mathrm{br},s}$ & 14.4$^{\circ}$, 6.24$^{\circ}$, 1.9$^{\circ}$, 1.37$^{\circ}$ \\
     \hline 
    \end{tabular}
    \caption{System parameter values}
    \label{Table 1}
\end{table}

The path gain parameter, $P$, in \eq{\ref{Eq: Path gain matrices for Hd and Hru}} is selected such that the average total channel power for a single user is 0dB for a baseline case where $M=64,N=100,\kd=\kru=0$ and $\Hbr{}$ is pure LOS. This gives $P=45$dB. The results for all simulations are averaged over $10^4$ user locations and their associated path gains. 

\begin{figure}[h]
	\centering
	\myincludegraphics{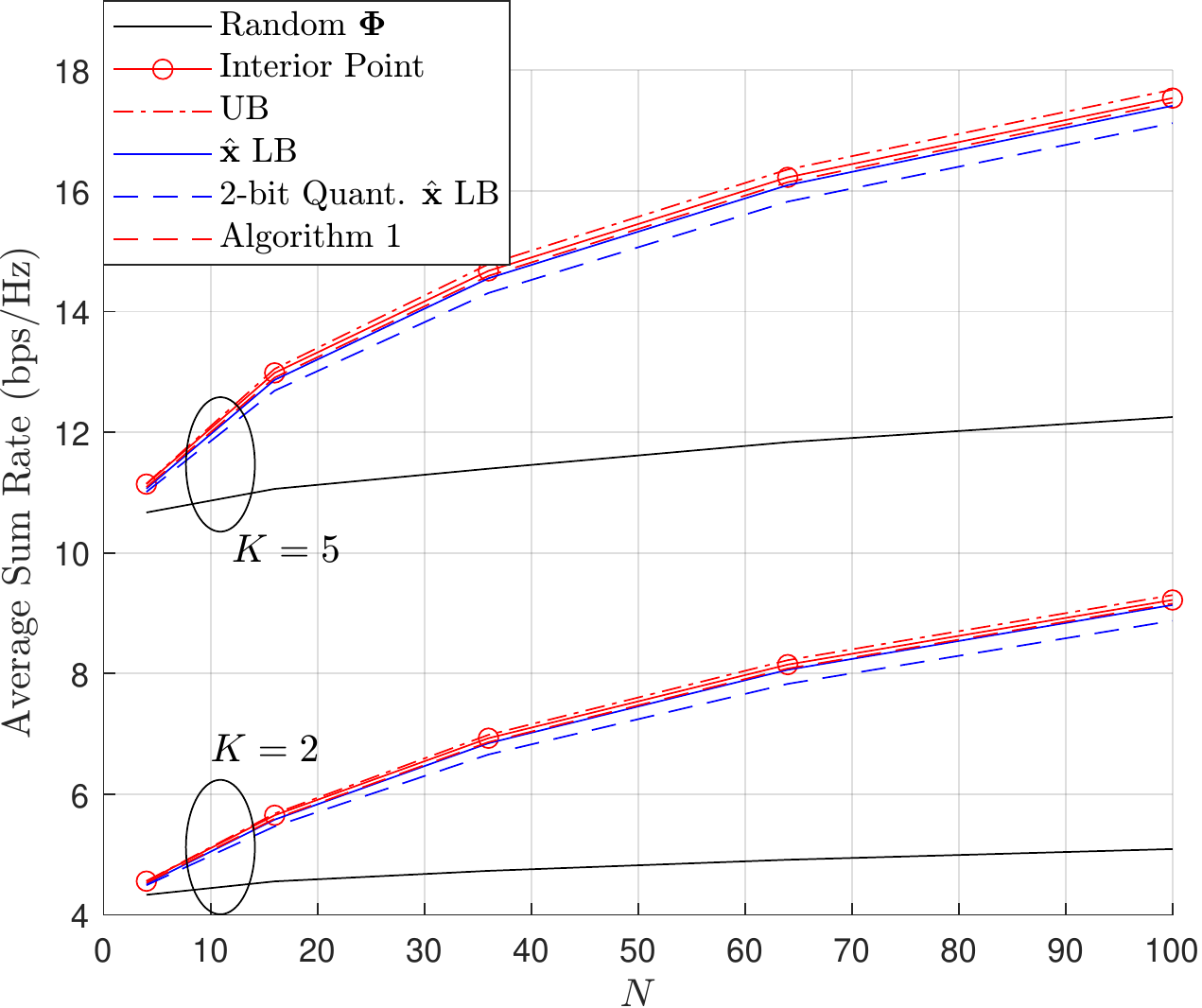}
	\caption{Average sum-rate bounds and approximations for varying $N$ and $\kru=\kd=1,\kbr=\infty$, $ K\in\{2,5\}$.}
	\label{Fig: Fig 2}
\end{figure}

First, in Fig.~\ref{Fig: Fig 2}, we demonstrate the effectiveness of the optimization techniques presented in Sec.~\ref{Sec: Sum-Rate Maximization} for varying RIS sizes. Here, we show the lower bound \eq{\ref{Eq: Closed form optimal RIS}}, a 2-bit quantization of the lower bound, the AO algorithm shown in Alg.~\ref{Algorithm 1} and the upper bound given in \eq{\ref{Eq: Upper bound on Sum-Rate}}. These expressions are computed for scenarios where $\kru=\kd=1$ and $\kbr=\infty$ to represent a pure LOS RIS-BS channel The number of UEs is $ K \in\{2,5\}$. We compare these results against two benchmark cases:
\begin{itemize}
    \item the optimal sum-rate computed by built-in numerical optimization software using the interior point algorithm; 
    \item the sum-rate  achieved by a random set of RIS phases selected from $\mathcal{U}(0,2\pi)$. 
\end{itemize}
As can be seen in Fig.~\ref{Fig: Fig 2}, the gain in sum-rate relative to random phases increases with $N$. The extremely simple upper and lower bounds derived via channel separation are shown to be very tight for all $N$. The simple AO approach gives results just below the interior point algorithm, possibly due to incomplete convergence or local maxima. However, the AO method remains efficient for large RIS sizes where run-time is a problem for the interior point algorithm. Note that quantizing the lower bound RIS design in \eqref{Eq: Closed form optimal RIS} provides a lower bound on the sum-rate achievable by a quantized RIS, which is demonstrated in Fig.~\ref{Fig: Fig 2} for 2-bit RIS phases.

\begin{figure}[h]
	\centering
	\myincludegraphics{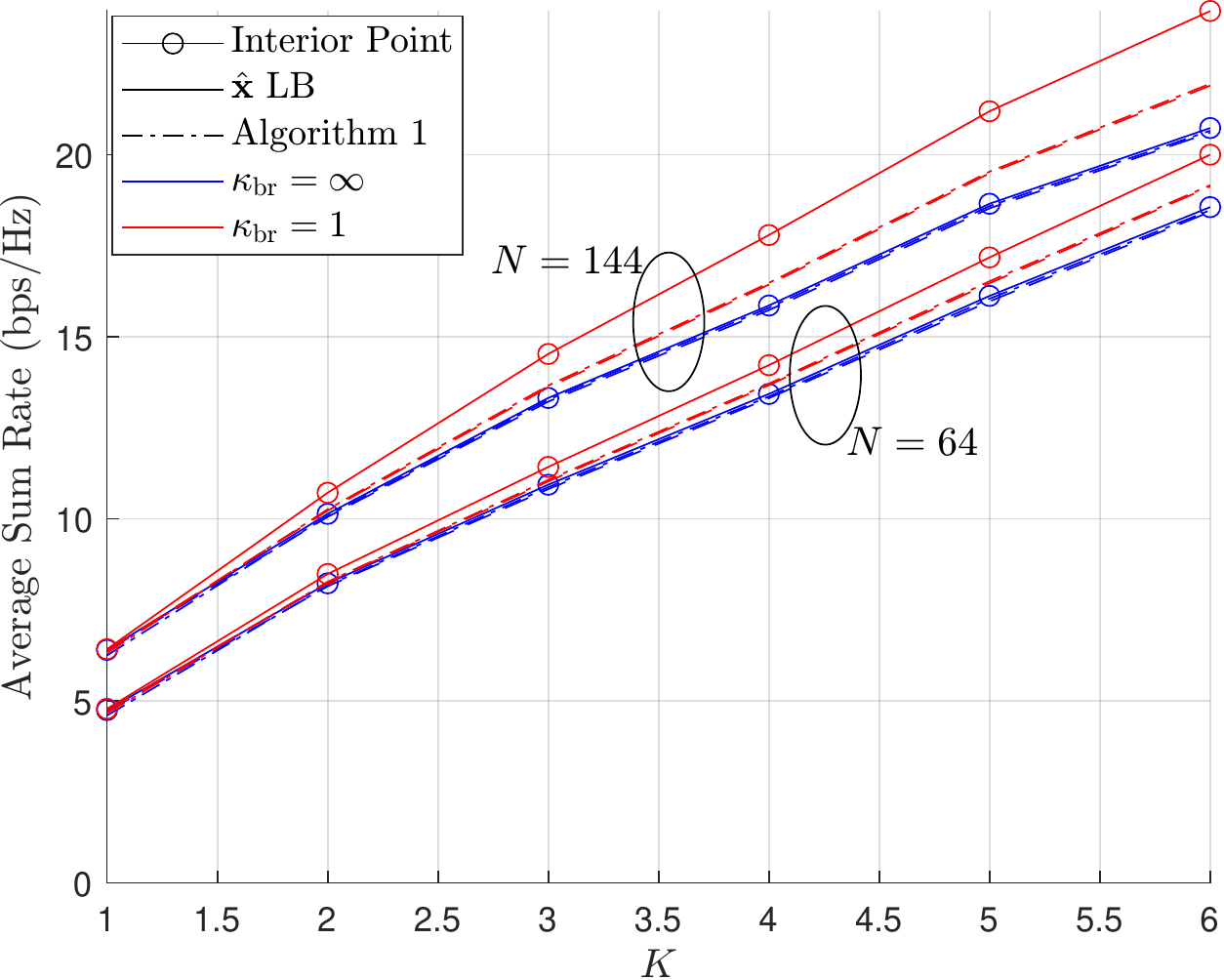}
	\caption{Average sum-rate using numerical optimzation, the AO algorithm and the lower bound for varying numbers of UEs, $N\in\{64,144\}$, $\kd=\kru=1$, $\kbr\in\{\infty,1\}$.}
	\label{Fig: Fig 3}
\end{figure}

As the lower bound in \eqref{Eq: Closed form optimal RIS} and the AO approach are based an actual RIS designs assuming pure LOS for the $\Hbr{}$ channel, applying these solutions to any other channel also results in lower bounds, although they may be less tight. To study the robustness of the lower bound and the AO algorithm, we compare the sum-rate against a varying number of UEs for systems where $\kd=\kru=1$, $\kbr \in \{\infty,1\}$, $N \in \{64,144\}$ and the results are shown in Fig.~\ref{Fig: Fig 3}.
As expected, since channel separation is designed for LOS RIS-BS channels, we see sub-optimal performance using the lower  bound and the AO algorithm for increasing UEs and RIS sizes while the pure LOS channels give extremely accurate results. However, even with $\kbr = 1$ (where the LOS and scattered powers are equal), the lower bounds are still reasonable. Hence, the simple bounds are still useful for RIS-BS channels which contain some scattering but have a dominant LOS component.

\begin{figure}[h]
	\centering
	\myincludegraphics{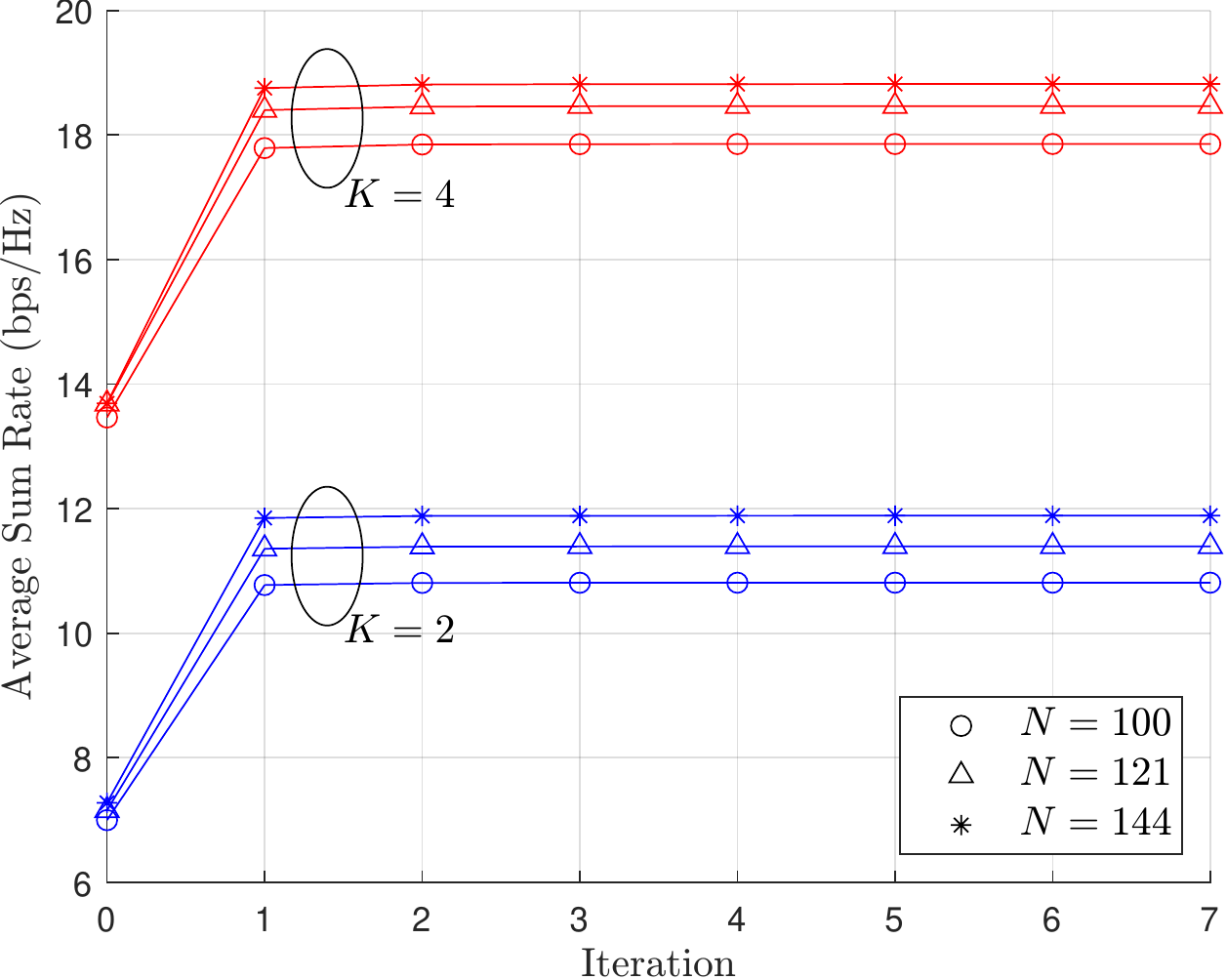}
	\caption{Convergence of AO Algorithm for systems with $K=2$ and $K=4$ UEs, both with $N \in \{100,121,144\}$ RIS sizes.}
	\label{Fig: Fig 4}
\end{figure}

In Fig.~\ref{Fig: Fig 4}, we demonstrate the average efficiency of the AO algorithm by plotting the average sum-rates achieved at various iterations of the algorithm. Results are presented for systems with $K=2$ and $K=4$ UEs both with $N \in \{100,121,144\}$. The channel parameters are $\kd=\kru=1$ and $\kru=\infty$.
Even for large RIS sizes and increasing numbers of UEs, Fig.~\ref{Fig: Fig 3} shows that the algorithm converges very quickly. This is the property which enables rapid AO results to be obtained for large $N$, while numerical optimization becomes extremely slow.

\section{Conclusion}\label{Sec: Conclusion}
In this paper, we have presented a \textit{channel separation} technique which allows for a new understanding on the effects of RIS phases on the sum-rate. Specifically, channel separation creates an equivalent channel matrix separated into two parts; one part is independent of the RIS and another part consists of a single row directly impacted by the RIS. Leveraging this technique, we derive extremely simple upper and lower bounds on the optimal sum-rate. In addition, we propose a low-complexity AO algorithm to obtain sub-optimal sum-rate results. Numerical results demonstrate the effectiveness of the presented techniques. Despite their simplicity, the bounds are shown to be very tight and the AO algorithm converges very quickly even for systems with large RIS sizes. In scenarios where hardware limitations are present, quantizing our proposed solutions leads to tight lower bounds on the sum-rate achievable with quantized RIS designs and the resulting quantization degradation is shown to be minor. Although channel separation is designed for scenarios where the RIS-BS channel is LOS, the resulting lower bounds are demonstrated to be robust when the RIS-BS channel contains a weaker scattered component.

\bibliographystyle{IEEEtran}
\bibliography{main.bib}

\begin{thebibliography}{1}
\providecommand{\url}[1]{#1}
\csname url@samestyle\endcsname
\providecommand{\newblock}{\relax}
\providecommand{\bibinfo}[2]{#2}
\providecommand{\BIBentrySTDinterwordspacing}{\spaceskip=0pt\relax}
\providecommand{\BIBentryALTinterwordstretchfactor}{4}
\providecommand{\BIBentryALTinterwordspacing}{\spaceskip=\fontdimen2\font plus
\BIBentryALTinterwordstretchfactor\fontdimen3\font minus
  \fontdimen4\font\relax}
\providecommand{\BIBforeignlanguage}[2]{{%
\expandafter\ifx\csname l@#1\endcsname\relax
\typeout{** WARNING: IEEEtran.bst: No hyphenation pattern has been}%
\typeout{** loaded for the language `#1'. Using the pattern for}%
\typeout{** the default language instead.}%
\else
\language=\csname l@#1\endcsname
\fi
#2}}
\providecommand{\BIBdecl}{\relax}
\BIBdecl

\bibitem{8796365}
E.~Basar \emph{et~al.}, ``Wireless communications through reconfigurable
  intelligent surfaces,'' \emph{IEEE Access}, vol.~7, pp. 116\,753--116\,773,
  2019.

\bibitem{8741198}
C.~Huang \emph{et~al.}, ``Reconfigurable intelligent surfaces for energy
  efficiency in wireless communication,'' \emph{{IEEE} Trans. Wireless
  Commun.}, vol.~18, no.~8, pp. 4157--4170, 2019.

\bibitem{9090356}
C.~Pan \emph{et~al.}, ``Multicell {MIMO} communications relying on intelligent
  reflecting surfaces,'' \emph{{IEEE} Trans. Wireless Commun.}, vol.~19, no.~8,
  pp. 5218--5233, 2020.

\bibitem{9110889}
B.~Di \emph{et~al.}, ``Hybrid beamforming for reconfigurable intelligent
  surface based multi-user communications: Achievable rates with limited
  discrete phase shifts,'' \emph{{IEEE} J. Sel. Areas Commun.}, vol.~38, no.~8,
  pp. 1809--1822, Aug 2020.

\bibitem{9286726}
Y.~Zhang \emph{et~al.}, ``Reconfigurable intelligent surface aided cell-free
  {MIMO} communications,'' \emph{{IEEE} Wireless Commun. Lett.}, vol.~10,
  no.~4, pp. 775--779, April 2021.

\bibitem{9203956}
M.~Zeng \emph{et~al.}, ``Sum rate maximization for {IRS}-assisted uplink
  {NOMA},'' \emph{{IEEE} Commun. Lett.}, vol.~25, no.~1, pp. 234--238, 2021.

\bibitem{8812955}
C.~L. {Miller} \emph{et~al.}, ``Analytical framework for full-dimensional
  massive {MIMO} with ray-based channels,'' \emph{{IEEE} J. Sel. Topics Signal
  Process.}, vol.~13, no.~5, pp. 1181--1195, 2019.

\bibitem{9066923}
Q.-U.-A. Nadeem \emph{et~al.}, ``Asymptotic max-min {SINR} analysis of
  reconfigurable intelligent surface assisted {MISO} systems,'' \emph{{IEEE}
  Trans. Wireless Commun.}, vol.~19, no.~12, pp. 7748--7764, 2020.

\bibitem{dush}
D.~A. Basnayaka \emph{et~al.}, ``Ergodic sum capacity of macrodiversity {MIMO}
  systems in flat {R}ayleigh fading,'' \emph{{IEEE} Trans. Inf. Theory},
  vol.~59, no.~9, pp. 5257--5270, 2013.

\end{thebibliography}
\end{document}